# HRDBMS: Combining the Best of Modern and Traditional Relational Databases

Jason Arnold, Boris Glavic, Ioan Raicu
Department of Computer Science
IIT
Chicago, USA
jarnold6@hawk.iit.edu, bglavic@iit.edu, iraicu@cs.iit.edu

*Abstract*— HRDBMS is a novel distributed relational database that uses a hybrid model combining the best of traditional distributed relational databases and Big Data analytics platforms such as Hive. This allows HRDBMS to leverage years worth of research regarding query optimization, while also taking advantage of the scalability of Big Data platforms. The system uses an execution framework that is tailored for relational processing, thus addressing some of the performance challenges of running SQL on top of platforms such as MapReduce and Spark. These include excessive materialization of intermediate results, lack of a global cost-based optimization, unnecessary sorting, lack of index support, no statistics, no support for DML and ACID, and excessive communication caused by the rigid communication patterns enforced by these platforms.

## I. INTRODUCTION

The increasing scale of data to be processed for analytics has brought traditional database systems that scale only to a few nodes to their limits. Massively Parallel Processing (MPP) databases provide better scale out by parallelizing query processing across multiple processors and nodes using a shared-nothing architecture. While this proved to be effective at very small numbers of nodes, this approach did not scale to even medium-sized clusters. Examples of this type of approach include DB2 from IBM. DB2 is limited to clusters of 1000 nodes or less, but in practice few customers go above 10s of nodes. With extensive tuning, IBM has published some benchmarks using clusters of up to 256 nodes. [1] Another example of this type of solution is Netezza which is a database appliance sold by IBM. The primary difference between Netezza and DB2 is that Netezza runs on customized hardware, which includes offloading some processing to FPGA cards. The largest Netezza model today has 58 nodes. [2] Likewise, Teradata and Greenplum, two popular MPP databases for large analytics environments, have cluster size limits of 1024 [3] and 1000 [4] respectively.

Main-memory database systems based on columnar storage proved to be more efficient than their row-based counterparts for analytical queries and better exploit current hardware. However, they suffer the same scalability issues as standard MPP databases. For example, HANA is an extremely popular columnar in-memory database. The largest HANA cluster that has been reported is 56 nodes [5]. The reality is that while many of these databases have system limits of around 1k nodes, they fail to scale well even to 100s of nodes.

Recently, a new class of SQL engines has been built on top of Big Data platforms such as MapReduce and Spark. Examples of such systems are Hive [6], Spark SQL [7], Azure SQL[8], Dremel [9], and many others. While these approaches provide better scale out, their performance per node is poor, because query execution is restricted by the programming and execution model of the underlying platform. While the performance per node is poor, the platforms that these databases run on have been shown to scale to very large clusters. For example, Yahoo! runs a 4.5k node Hadoop cluster. [10]

*HRDBMS*, the system we present in this work, is an architectural hybrid that combines the scale out of Big Data analytics platforms with advanced query execution and optimization of traditional databases. The goal of HRDBMS is to create a database capable of scaling well past the 1k node limit, while achieving per-node performance on par with traditional MPP relational databases.

## II. RELATED WORK

Traditional MPP databases, in-memory databases, and Big Data platform databases all have a few other shortcomings as well. HRDBMS attempts to address these too. For example, many of these modern analytics databases do not support full Data Manipulation Language (DML) operations. DML is a subset of standard SQL that includes the INSERT, UPDATE, and DELETE statements. Many of these databases also do not support full transactional isolation and consistency (known as ACID in the database world). The table below summarizes the main shortcomings from section 1 plus these for some of the most popular analytics databases in the enterprise world.

|  | Supports Large Clusters | Full DML | Full ACID | Good performance data > mem | Standard syntax for basic SQL | Commodity Hardware |
|---|---|---|---|---|---|---|
| DB2 |  | X | X | X | X | X |
| Netezza |  | X | X | X | X |  |
| Teradata |  | X | X | X | X |  |
| Greenplum |  | X | X | X | X | X |
| HANA |  | X | X |  | X | X |
| Hive | X |  |  | X |  | X |
| Spark | X |  |  |  | X | X |

There's also several database systems in the research world that are trying to address similar shortcomings.

Most of these databases are only semi-relational and therefore frequently do not use standard SQL as their query language. As pointed out by others, many people believe that large scale systems and relational models are mutually exclusive. [8]

For example, Spanner [11] from Google is a semi-relational database designed to scale to millions of nodes. Only thousands of nodes at most would typically be used to support an individual query, but Spanner provides the ability to isolate different sets of data on different nodes and provides the ability to do a large number of replicas on a global scale. The main focus of Spanner though is cross data center replication of data for availability in the event of data center loss. They forego some of the things we expect from relational databases in order to achieve their goal. For example, their query language is SQL-like but is different enough to pose problems for using Spanner with standard tooling. They also use a semi-relational schema that is a cross between key/value pairs and standard relational schemas. It's basically a relational view of the data built on top of a key/value store. But, it's also a hierarchical data structure in that all tables are defined in terms of where they sit in the hierarchy that defines their relationship to other tables. The authors state that this is for performance since the hierarchy automatically defines what data should be co-located. However, many real world data models can't be expressed in terms of simple hierarchies such as these.

Since the underlying storage engine is key/value store based, primary indexes are not needed for efficient lookup by primary key. However, it does make it challenging to implement secondary index support, and in fact, Spanner does not offer secondary index support today. The authors do point out that some applications that use Spanner have built application-side processes to mimic secondary indexes to get the performance they need. HRDBMS, on the other hand, allows you to create whatever secondary indexes you need.

F1 [12] is another database from Google that is built on top of Spanner. F1 addresses some of the shortcomings of Spanner such as lack of secondary indexes and lack of standard SQL support. However, it does this by basically using Spanner as a storage engine. F1 implements its own optimizer and uses its own cluster of servers to do the actual query processing. Because it relies on Spanner as the underlying storage engine only, it foregoes the possibility of taking advantage of existing co-locality in the data storage. The authors state that F1 scales well to hundreds of servers, so it does not appear to solve the scalability problem that traditional MPP databases face.

Dremel [9], yet another database from Google is designed for analytics across thousands of cpus. But, it does lack DML support. The data is read-only. Dremel also uses a nested data model instead of a standard relational model and only offers a SQL-like query language.

None of the previously mentioned databases have published any performance data for TPC-H or other standard analytics workloads, so it's difficult to ascertain where they lie in the performance spectrum.

Azure SQL (SQL Azure / Cloud SQL Server) [8] is much more similar to HRDBMS in that their major focus is on supporting a standard relational model and standard SQL. They also are focused on supporting ACID in an analytics environment, although they only offer ACID guarantees for certain types of transactions today.

Related to this area of research is the topic of key/value stores, such as ZHT. [13] As previously mentioned Spanner is a semi-relational database built on top of a key/value store. Much more progress has been made in building scalable key/value stores, so it is a reasonable approach. The problem is that key/value stores in and of themselves do not provide the features needed for analytics. As mentioned with Spanner, the underlying key/value store makes it difficult to implement secondary indexes. When F1 addressed these and other issues by building on top of Spanner it lost much of the scalability that Spanner had. Furthermore, key/value stores in and of themselves do not offer a query language capable of expressing complex analytics queries. They also generally do not have optimizers, since the means to respond to certain data requests is usually obvious. Complex analytics queries on the other hand rely on a very expressive query language and heavy use of optimization.

In the following sections, we will cover how HRDBMS combines the best of both the traditional relational database world and the Big Data platform database world. The key contributions of each area are summarized here.

- Traditional relational databases
    - Query optimization
    - Pipelining
    - Use of statistics to optimize
    - Ability to perform operations externally if not enough memory is available
- Big Data platform databases
    - Map/Combine/Reduce/Shuffle computation model

- o Phase combination (from Spark) to reduce communications

HRDBMS combines the best of both of these worlds by:

- Starting with a standard operator tree representation of the query
- Applying standard relational optimizations, using statistics
- Converting to a Map/Combine/Shuffle/Reduce representation
- Pipelining in the execution engine
- Using a custom non-blocking shuffle in the execution engine
- Allowing operations to spill to disk (external) if need be
- Using a more flexible map/reduce model that allows chaining of map and combine phases without intervening reduce phases

The key novel contributions of HRDBMS are:

- A database optimizer that is capable of performing traditional relational database optimizations but then converting that to an optimized MapReduce-like job.
- A custom MapReduce-like execution engine that is tailored for and tuned for database operations and relational algebra.
- A more flexible shuffle that reduces the network communications burden for large clusters.

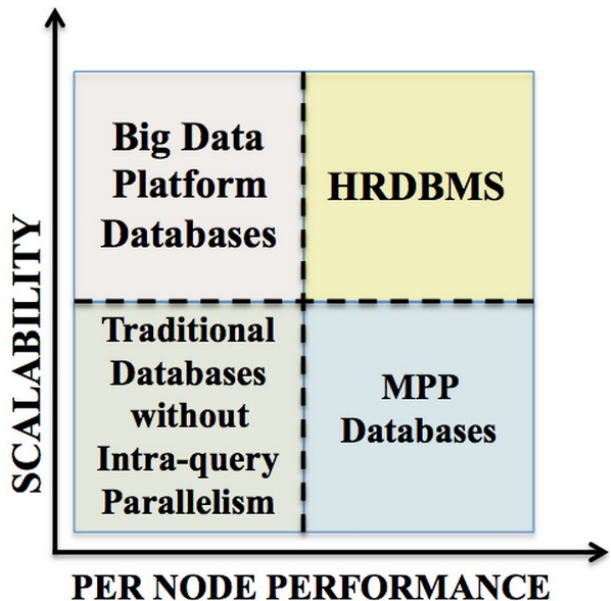

The rest of this paper is organized as follows. Section 3 defines key relational database concepts that are required for understanding the rest of the paper. Section 4 describes the general architecture of an HRDBMS database cluster. Section 5 provides details on the custom execution engine. Section 6 explains the workings of the optimizer. Section 7 shows experimental results to date. Finally, section 8 discusses some of our thoughts for future research.

### III. KEY RELATIONAL DATABASE CONCEPTS

Throughout the rest of this paper we will be discussing many concepts that are taken for granted in the database world that may be foreign to some of the readers. This section provides an introduction to the key concepts that are required to understand the rest of the paper.

SQL – SQL is the language used to communicate commands with a database server. It consists both of SELECT statements, which represent queries, as well as INSERT/UPDATE/DELETE statements, which modify the data. The statements that modify data are typically called Data Manipulation Language (DML) statements.

Tables – Data in a relational database is organized and stored in tables. Each table has a unique name. The data in the tables is organized into rows and columns. A row represents a record. A row is made of multiple columns, each with a pre-defined data type. SQL statements are only valid if they perform valid operations against valid data types. For example, we can't add columns A and B if they are not both numeric. We can't perform a substring on a column unless it is character data.

Indexes – Many queries contain predicates (filter conditions) that can be logically ANDed or ORed together in the WHERE clause of a SELECT statement. If there are few enough rows that will satisfy a certain predicate (selectivity in DB jargon), it may be beneficial to build an index to support that predicate. An index is an ordered tree structure that allows the database to quickly determine which rows in a table will satisfy a predicate instead of reading (scanning) the whole table. The index will return a set of row ids (RIDs) and then only these rows are fetched.

Query optimization – When a SQL query enters the system, it is parsed and converted to a tree structure of relational operators that will produce the correct result. The goal of an optimizer is to use relational algebra equalities to reorder and restructure the tree of relational operators so that the same end result is achieved in a much more efficient manner. Query optimizers typically improve performance by several orders of magnitude over the naïve representation that comes out of the parser. The decision of whether to use an index or do a table scan is an example of one of the types of decisions that the optimizer needs to make. Query optimization is a notoriously hard problem with a huge search space, so frequently heuristics are used to limit the number of possibilities that have to be considered.

Join – In a relational database, a join is the union of columns from 2 different tables into a single row based on some predicate that spans the 2 tables. For example, assume we have the following 2 tables.

| Student_ID | Student_Name | Advisor_ID |
|---|---|---|
| 1 | Student_1 | 1 |
| 2 | Student_2 | 8 |
| 4 | Student_4 | 2 |
| 5 | Student_5 | 3 |
| 7 | Student_7 | 3 |
| 9 | Student_9 | 1 |
| 10 | Student_10 | 3 |

| Advisor_ID | Advisor_Name |
|---|---|
| 1 | Advisor 1 |
| 3 | Advisor 3 |
| 5 | Advisor 5 |

A join on Advisor_ID would give us the following result.

| Student_Name | Advisor_Name |
|---|---|
| Student_1 | Advisor 1 |
| Student_5 | Advisor 3 |
| Student_7 | Advisor 3 |
| Student_9 | Advisor 1 |
| Student_10 | Advisor 3 |

[14]

Distributed Join – When a relational database runs on a cluster, it has to perform joins in a distributed manner. In other words, it wants to be able to execute joins in parallel across all of the nodes of the cluster. The most common method for doing this is to redistribute all of the data across the cluster by hashing on the join key (in the case of equality predicates). This guarantees that the rows that need to be joined together from the left and right side of the join will be on the same node. This then allows the join to be computed as numerous parallel joins on each node followed by the union of all of the results.

Semi-join – A semi-join is like a join except that none of the data from the right-hand part of the join is included in the output result. Instead, only rows that have a match (based on the join predicate) in the right-hand side are included in the output result. An anti-join is the opposite of this. Only rows from the left-hand side that do not have a match on the right-hand side (based on the join predicate) are included in the output result. Therefore, semi-joins are frequently referred to as existence tests (in fact they correspond directly to the EXISTS keyword in SQL), and anti-joins are referred to as non-existence tests (NOT EXISTS in SQL).

Two Phase Commit (2PC / XA) – Two phase commit is a standard protocol for ensuring that transactions that span multiple machines all commit together on every machine or all rollback together on every machine. HRDBMS uses a custom variant of the 2PC protocol that uses hierarchical network communications and aggregated responses to make the protocol faster for larger clusters.

IV. HIGH LEVEL ARCHITECTURE

An HRDBMS cluster is divided into *coordinator* and *worker* nodes. Coordinator nodes store metadata and are responsible for query optimization. Worker nodes are used for user data storage and the majority of query execution work. Clients submit SQL commands to a coordinator node. The coordinator node parses and optimizes incoming queries using data distribution statistics and partitioning information stored in system tables which are replicated across coordinators. Like traditional relational systems, HRDBMS uses a cost-based optimizer (Section VI). The output of the optimizer is a query workflow, which is submitted for execution to the cluster. The workflow executes across the worker nodes on the custom MapReduce based platform (Section V). Query results are eventually sent back to the coordinator that planned the query. Final sorts or aggregations may occur on the coordinator, if they are cheap enough. The coordinator then forwards the result set to the client.

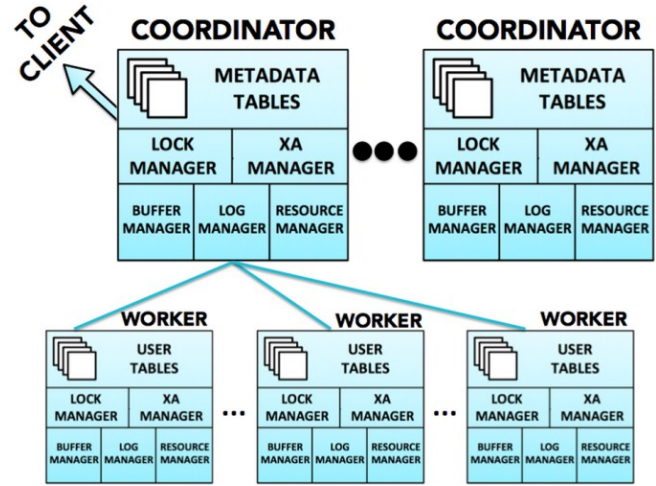

When a query is received by a coordinator node, it first has to go through parsing. This is no different than any of the systems previously mentioned. However, differences start coming up almost immediately after that. First of all after the query is parsed, it is converted from an abstract syntax tree to a tree of relational operators (select, project, union, join, etc). This is a standard representation for queries in traditional relational databases. Next, statistics are used to apply standard relational optimizations to the operator tree. Big Data databases like Spark SQL and Hive generally do not have statistics other than total data size. In contrast, traditional relational operator tree optimizations rely heavily on statistics (cardinality stats, distribution stats, and more). The statistics are stored in system metadata tables that reside on all coordinator nodes. The data required to parse and verify that queries are valid is also stored in these system metadata tables. These system metadata tables are no different than user data tables, other than the fact that they (and they alone) reside on the coordinator nodes. They are queryable, but not updatable, by users.

From there, the query is converted to a workflow format and additional optimizations are carried out. These will be covered in detail in section V.

Finally, the coordinator submits the workflow to the execution engine and return results to the client. Since HRDBMS does support transactions, this workflow may be one of many that is part of the same transaction. The coordinator is also responsible for tracking and managing transactions until they either commit or rollback. This requires the coordinator to have a number of other services available to it. This includes logging services (both for changes to system metadata tables as well as a two phase commit log). It has XA (2PC) coordinator services, which uses a custom hierarchical two phase commit protocol. It has locking services. For example, if someone is currently loading a table, you shouldn't be able to query against it until the load finishes. It has buffering services, so that pages need not always be read from disk. This is another item noticeably lacking in Big Data databases. Lastly, the coordinator nodes have full blown execution engine capabilities, although they are normally only used for executing single node workflows that handle metadata table queries.

The worker nodes are very similar in architecture to the coordinator nodes. They contain the logging services necessary for handling transactions against the user data. They contain the locking services necessary for handling those transactions. They also contain the buffering services to avoid as many reads to disk. However instead of being able to act as XA coordinators, they have the ability to act as XA slaves, and XA intermediaries. XA intermediaries are used to forward XA messages as part of the custom XA implementation I have written and to aggregate XA responses. They also contain a full-blown execution engine, although the worker execution engines will typically be much more active than the coordinator execution engines since they handle the majority of query processing.

HRDBMS has full DML support unlike Hive and Spark SQL. These restrictions in Hive and Spark SQL primarily come about because they rely on HDFS, the Hadoop filesystem. HDFS only supports append operations to files. It does not support rewriting existing contents. This makes it very challenging to support full INSERT/UPDATE/DELETE capabilities. Since HRDBMS does not use HDFS, it does not suffer these shortcomings.

As mentioned previously, HRDBMS also deploys a handful of common services on each node which allows it to support transactions. Hive and Spark SQL rely solely on the underlying workflow engine and do not have specialized components deployed to each node. Thus they cannot handle transactions.

HRDBMS is also designed from the ground up to handle large data volumes and does not rely on being able to fit data in memory. All possible workflow actions are also designed to be able to run in memory or externally depending on the amount of memory available versus the amount required.

HRDBMS is not fully ANSI SQL 92 compliant, but all of the basic SQL constructs that are used in the majority of queries use standard SQL 92 syntax. As previously mentioned, many of the other databases in this space do not support standard SQL syntax.

Lastly, HRDBMS can run on any type of hardware. It is written in Java and is easily portable to any system capable of supporting a Linux-like filesystem. Additionally, HRDBMS can offload some of its processing to GPUs, if they are present.

V. HRDBMS'S DISTRIBUTED EXECUTION ENGINE

While also executing queries as tasks that run independently on worker nodes like Big Data platforms do, HRDMBS pipelines operations (even across nodes), enforces data locality, makes use of index structures, and opportunistically materializes intermediate results to disk if they do not fit into main memory. In HRDBMS there is no separation of jobs into phases. Operators of a query immediately start consuming their inputs when they become available.

Some of the major enhancements of the custom execution engine are summarized in the following sub-sections.

**Pipelined Execution.** Pipelining is a standard practice in many traditional relational databases that greatly helps improve performance [15]. Basically, a call to retrieve the next record from one task produces a call to retrieve the next record out of the previous task. The necessary transformations on that record are performed and the record is returned. Pipelining such as this is frequently many levels deep in relational databases, as long as a blocking operation does not intervene. In fact, HRDBMS may choose to sort buckets in parallel and then mergesort even when a sort fits in memory because only the sorting of the buckets blocks, while the mergesort does not.

A major contributor to the poor performance of Hive is excessive materialization. A query may require multiple map-reduce phases and during each phase intermediate results are written to disk twice – once to local disk at the end of the map phase and once to HDFS at the end of the reduce phase. Systems such as Spark address this problem, but still divide jobs into phases that execute sequentially. In HRDBMS, we forego the use of phases altogether and string a series of tasks together with shuffles in between them to redistribute data across nodes. These tasks can perform any set of relational operators, including aggregation. A task can start processing data before a previous task, producing its inputs, has finished execution. In fact for some queries, all of the tasks involved in the query may immediately begin to do work when query execution starts. This is made possible by pipelining and other innovations such as the non-blocking shuffle covered next.

**Non-blocking Shuffle.** In Hadoop and Spark, a shuffle operation redistributes records across the cluster. In Hadoop a shuffle is always a blocking operation since it sorts the records it is shuffling. This is because Reducer tasks expect the data to be received in sorted order within each key. Almost all relational database operations can be performed without this requirement. Therefore, we can eliminate this overhead, except where absolutely required. Spark can sometimes skip the sort step, but it still finishes processing its entire input before it starts sending data. Spark will still perform a sort during the shuffle if the shuffle will be followed by an aggregation operation. This is unnecessary as aggregation operations can be performed equally as efficiently using hashing as long as your aggregation operation can spill hash buckets to disk. Even when not sorting, Spark still uses a blocking shuffle since it hashes (and spills to disk) all of the records to be shuffled before sending any data out. This is so that it can limit the number of socket connections that it needs to have open at any given time when sending data. Basically it opens one hash bucket at a time and only needs to have open sockets for the nodes/keys that bucket contains. Then it moves on to the next hash bucket.

HRDBMS implements a non-blocking shuffle operation that partitions tuples across nodes using hashing. It does this without sorting and without spilling to disk. Tuples with the same hash value are guaranteed to be sent to the same node, but the shuffle does not guarantee any sort order within the set of tuples sent to a node. This does mean that HRDBMS is sending over multiple sockets at the same time. However, HRDBMS limits the amount of network overhead using a different approach that we call "hierarchical shuffle". When the number of other nodes that a node must communicate with becomes large, hierarchical shuffle kicks in. In a hierarchical shuffle, each node only communicates with a set of neighbors, which is a subset of the cluster. These neighbors can then forward data on to yet other nodes if needed. The hierarchical shuffle gives us hard limits on the number of socket connections that a shuffle will concurrently open, while still giving us the ability to shuffle in a non-blocking manner without the need to externalize to disk.

**Intra Task Parallelism.** Tasks within one node are inherently parallel. Each task is defined in terms of the I/O and relational operations it will perform. HRDBMS is designed such that it is easy to construct tasks which perform different relational operations in parallel, do I/O in parallel, and even use parallelism within the execution of a single relational operator. For example, table data is not only partitioned across nodes, but also partitioned across disk drives on each node. A separate I/O thread is assigned to each disk. This means that if a table needs to be scanned on some worker node, all disks are scanned in parallel, and any filter conditions that can be pushed down to the scan level are applied to the data in parallel. Furthermore, other complex operations such as joins, sorts, and aggregation are inherently performed using parallelization within each task on each worker node.

**External Operations**. All relational operations that are pushed into tasks in the query workflow have the ability to be performed in memory or externally, spilling data to disk as needed. This decision is made at runtime based on the current available memory and statistics which predict the amount of memory that will be required. Contrast this with Hadoop and Spark, which sometimes force data to be materialized for certain operations, although they may have been able to be performed in memory. Also, there are cases where Hive or Spark just fails if something can't fit in memory, whereas HRDBMS will just choose to perform that operation externally.

**Storage.** Instead of using a distributed file system such as HDFS, we store data directly on the local filesystem of worker nodes. Tables may be hash or range partitioned or duplicated across all nodes, in the case of small tables. The coordinator nodes track how table data is spread across worker nodes and uses this information during optimization. Data locality is enforced. This means that data is always read on the worker node where it resides. Contrast this with Hive or Spark SQL, where the database can only state this as a preference to the underlying execution engine. This preference may or may not be honored. This can result in non-local reads and destruction of any pre-existing co-location that may exist.

**Predicate Cache**. HRDBMS uses a predicate cache to further speed up table scan operations. If HRDBMS has previously seen a predicate and determined that a certain page has no rows that qualify for the predicate, this information is cached. If we then see the same or a stronger predicate, we don't have to even read certain pages.

> *Cache says: Page 5, no rows qualify for A > 5*
> *We know: Page 5, no rows qualify for A > 10 is also true*

The information in this cache is selectively invalidated if a page changes. I am aware of some databases that pre-calculate some of this information on selected columns ahead of time, with the assumption that deletes and updates will be rare and inserts will be appended. This data will then be recalculated when a table reorganization occurs. I am not aware of any database that calculates, caches, invalidates, and recalculates this information at runtime without reorganization.

**Hash Joins**. HRDBMS exclusively uses hash joins for any join operations that have at least 1 equality join condition. Given the importance of hash joins in HRDBMS, it was important to make them perform well. Therefore HRDBMS has a rather sophisticated hash join implementation. First of all hash joins can be completely in-memory, partially in-memory, or fully external. Partial in-memory hash joins are like hybrid hash joins where the first bucket is never written to disk. HRDBMS extends this by allowing the first n buckets to never be written to disk. The value of n is dynamically determined at runtime based on current memory availability and statistical information that is included in the workflow.

Secondly, data is stored in a compressed format in the in-memory hash tables. This allows us to fit more rows in memory and/or use less memory. This is important for performance for 2 reasons. First of all, it obviously has the potential to reduce the amount of data that must be written to and read from disk (by fitting more in memory). But, since HRDBMS is written in Java, it also allows us to perform fully in memory hash joins using less memory than would be required otherwise. This reduces the burden on the Java garbage collector and greatly improves performance. Testing shows us that the biggest impediment to further performance gains is Java garbage collection overhead.

Lastly, the hash join implementation makes heavy use of bloom filters. Many other databases use bloom filters for semi-joins [16]. HRDBMS uses bloom filters for hash joins, semi-joins (existence), and anti-joins (non-existence) since all three operations are implemented using the same hash join code. Basically both sides of the join compute a bloom filter as they are reading. When one side finishes reading its input data, it shares its bloom filter with the other side. That bloom filter is then applied to any further data read by that side.

**Buffer Manager**. The buffer manager in HRDBMS is also rather complex given the impact that buffering can have. The fact that HRDBMS even has a buffer manager is one of the major reasons why it so greatly outperforms Hive, as will be shown in section VII. The buffer manager caches both table and index pages. When part of the executing workflow needs a page, it informs the buffer manager. This page request can be either synchronous (I need this page now) or asynchronous (I need this page soon, please go get it and I'll tell you when I need it). This allows the workflow to do other work while the I/O (if necessary) occurs. If the page is present in the cache (or in DB jargon, the bufferpool), no disk I/O needs to occur.

After trying many different traditional and non-traditional cache eviction policies, I determined that the best approach was a standard "clock" eviction policy with a twist. The twist is that when a table scan operation starts, it informs the buffer manager about the entire set of pages that it will read (this starts off as being the whole table). As it reads, it informs the buffer manager about its progress, and the buffer manager updates its list of pages that are still going to be requested in the near future. When the buffer manager needs to evict a page to make room for a new one, it uses the standard clock eviction policy with the caveat that it attempts to find a page that is not currently registered as "still being needed" by any currently executing table scan. If the clock pointer goes around in a full circle without finding a page to evict (all pages are still needed), then the buffer manager falls back to a traditional clock eviction policy.

Lastly, the initial size of the bufferpool is specified as a database parameter, but the buffer manager has the ability to dynamically grow the bufferpool when needed to avoid bufferpool exhaustion errors. It also has the ability to dynamically shrink the bufferpool to free up memory for other purposes and reduce the burden on the Java garbage collector.

## VI. THE HRDBMS OPTIMIZER

The optimizer can be summarized as follows:

- Phase 1 – traditional relational optimizations
- Phase 2 – conversion to MapReduce-like workflow
- Phase 3 – basic workflow optimizations
- Phase 4 – advanced workflow optimizations
- Phase 5 – indexes and miscellaneous optimizations

**Phase 1**. HRDBMS takes advantage of decades of research in relational query optimization to build efficient execution plans. Query planning starts out very similar to query planning for a traditional relational database. Standard transformations are applied. This includes selection pushdown, where rows are eliminated as early as possible if it can be determined that they won't be needed upstream. We also do projection pushdown, where columns are eliminated as early as possible if they won't be needed later. These are not novel in and of themselves, and are well-documented elsewhere. What is novel is the ability to apply these types of optimizations in a Big Data type of platform.

The only other standard optimization worth mentioning is join enumeration. In a standard relational database, join enumeration is the process of figuring the best order for performing joins when queries contain more than 2 tables. This single step of the optimizer has more impact on the performance of the query than any other step. There are several approaches to join enumeration, but they all share a common goal of trying to figure out a way of processing the least amount of data while still generating the correct result. As the number of tables involved in the query goes up, the number of possible ways that the joins can be done goes up dramatically. The costs of these different join possibilities are usually several orders of magnitude in difference. All traditional relational databases rely on statistics about the data to determine a good join order. Since Big Data databases like Hive and Spark SQL don't have statistics information (they only know the total size of each table), they do a very poor job of figuring out a good join order.

There are many different algorithms developed over the years for doing join enumeration in a traditional relational database. HRDBMS uses one of the simpler traditional join enumeration algorithms with a few changes. It's basically a greedy join enumeration [17]. HRDBMS uses statistics to estimate the number of rows that will be output by each join possibility. It then orders the joins in a manner that attempts to minimize the sum of the number of rows processed by each join. The optimizer sometimes will choose an option that does not minimize this value if it instead allows us to take advantage of pre-existing data co-locality and the impact to the sum is not that great.

The HRDBMS optimizer also generates all the transitive joins prior to doing join enumeration. This can lead to even more possible join orderings and even faster query execution.

*For example, if the user specifies in their query that A = B AND B = C, we also that A = C as well even though it wasn't explicitly stated in the query.*

**Phase 2**. After the standard relational optimizations are applied, they operator tree is converted to a Map/Shuffle/Reduce type of workflow by phase 2 of the optimizer. This is a very naïve translation from an operator tree to a workflow. It has all of the worker nodes read the necessary data from their local filesystems, send all of that data back to the coordinator node, and the coordinator node does all the query processing.

**Phase 3**. In phase 3 of the optimizer, we start to convert this into a much smarter workflow. We go through and figure out how all the various relational operations can be applied in parallel across the workers. For example, equi-joins can be performed in parallel as long as the data is distributed across workers based on the join key. Aggregations (count, min, max, sum ,and average) can be performed in parallel without any explicit partitioning. You just aggregate separately on each worker and then aggregate the aggregates again on the coordinator. You do have to convert average into separate sum and count operations for this to work. Likewise, sorts can be done in parallel by performing a final mergesort on the coordinator. There's corresponding rules for parallelization of all of the relational operators.

Now that we've figured out how the data needs to be partitioned at each step of the process, we build a workflow that has map phases on each worker that do what work they can with the current data partitioning, they then do a non-blocking shuffle to repartition the data in the manner needed by the next map phase. So, at this point we basically have a chain of map-shuffle-map-shuffle-… and finally everyone maps back to the coordinator where a reduce does final aggregations and sorts.

**Phase 4**. In Phase 4 of the optimizer, we attempt to make the workflow better yet. We now go through the workflow and track how the data is distributed in each map phase. We then remove any shuffle operations that can be removed. For example, if 1 map phase requires that the data is hash distributed by the value of column A (all rows with col A=x are on the same worker node) and the next map phase requires that the data is hash distributed by both column A (first) and column B (second), we can actually remove the shuffle operation since guaranteeing that all the common A values are on the same worker automatically implies that all the common (A,B) pairs are on the same worker. Any time that we are able to remove a shuffle operation, we now have 2 map phases that are directly connected to each other. These 2 map phases are then combined into a single map phase. This is similar to the phase combination process used by Spark.

**Phase 5**. In Phase 5 of the optimizer, we evaluate replacing table scans with index scans or index probes. If there is an index that contains all of the data that we need from the table and that index is smaller in number of pages than the table, we will scan the index instead of the table. If there are predicates to be applied to the data that comes from the table scans, statistics estimate that a very small number of rows will pass the predicate, and we have an index available to support the predicate, we will replace the table scan with an index probe. In an index probe, we position in the index based on the predicate and then scan a very small subset of the index to find rows that qualify. Instead of returning column data from the index probe, we return row identifiers (16 byte ID that unique represents a row in a table). We then go fetch only the necessary pages and rows from the tables by row ID.

Indexes in HRDBMS are represented both in memory and on disk as skip lists. Skip lists for in memory indexes is fairly common in the database world, but I am not aware of indexes as skip lists being stored on disk. Most databases store their indexes on disk as B+ trees. HRDBMS originally had a B+ tree implementation, but it had a very negative impact on load performance due to pages splits and other reasons that records had to be moved around. Replacing the B+ tree implementation with a (much simpler) skip list implementation greatly improved load performance and the difference in query performance was found to be statistically insignificant.

In our implementation of skip lists on disk we are almost guaranteed that if we follow a down pointer that we remain on the same page. If we follow a right pointer, we are almost guaranteed to go to a different page. If we use a promotion rate of 1/x, then the expected number of right pointers that have to be traversed at each level is x. This means that for an index probe:

Expected # page reads = $x \log_x n$

If we solve for x to minimize this number, we get:

$$\frac{\partial}{\partial x}(x \log_x(n)) = \frac{\log(n)(\log(x) - 1)}{\log^2(x)}$$

$$\frac{\log(n)(-1 + \log(x))}{\log^2(x)} = 0$$

$$x = e$$

So, we use a promotion rate of 1/e. Just as with B+ tree indexes, the closer you get to the top of the skip list, the more likely that the page will be found in the bufferpool.

All index and table data in HRDBMS is stored in 128KB pages. Every 3 pages makes up a 384KB superblock, which is LZ4-HC compressed by itself before being stored on disk. LZ4 was chosen as it has decent compression ratio and speed, but more importantly it has extremely fast decompression [18]. The decent compression ratio combined with the extremely fast decompression results in significant table/index scan performance improvements over not using compression.

Lastly, phase 5 finishes with a few optimizations that didn't fit well in any other phase. These miscellaneous steps are shown in the pseudocode below. Finally, the execution plan is composed and it is submitted to the HRDBMS execution framework.

The HRDBMS optimizer is over 11k lines of code (all of HRDBMS is over 150k lines), but the high-level pseudo code of the optimizer looks like this:

```
Optimizer Pseudocode
Phase 1
    1) Push down selects
    2) Join enumeration
    3) Products and selects combine to
    be joins
    4) Selects get pushed into table
    scans
```

```
    5) Push down projects
    6) Add projects for semi and anti
    joins
Phase 2
    1) Determine if we can skip any
    nodes or disks when reading
    table/index data
    2) Build naïve workflow
Phase 3
    1) Push down group by across join
    2) Implement hierarchical shuffle
    where needed
    3) Parallelize the relational
    operations across the cluster by
    stringing together map and shuffle
    phases
Phase 4
    1) Combine map phases by tracking
    data partitioning
    2) If we end up with a large sort at
    the end of the query on the
    coordinator, rewrite this as a
    parallel sort across workers with a
    mergesort on the coordinator
    3) Swap left/right children of hash
    joins so that the left side is
    always the higher cardinality
Phase 5
    1) Replace table scans with index
    probes where it makes sense
    2) Tell large hash group by's that
    they need to run externally (this is
    the only internal/external decision
    not made at runtime)
    3) Populate the workflow with
    statistics and cardinality
    information
    4) Parallel sorts across workers,
    followed by a mergesort and a limit
    on the coordinator get handled
    specially
    5) Replace table scans with index
    scans where it makes sense
    6) Figure out how to break the
    workflow up into pieces to be sent
    out such as to reduce the amount of
    data that has to be sent
```

Here is an example of a query that joins a couple tables together and does aggregation. This query will tell us how much money Canadian customers have spent. The picture below shows the initial operator tree when parsing is complete and the final workflow that comes out of the HRDBMS optimizer.

```
SELECT SUM(L_EXTENDEDPRICE)
    FROM LINEITEM,
        ORDERS,
        CUSTOMER,
        NATION
    WHERE O_ORDERKEY = L_ORDERKEY
    AND O_CUSTKEY = C_CUSTKEY
    AND C_NATIONKEY = N_NATIONKEY
    AND N_NAME = 'CANADA'
```

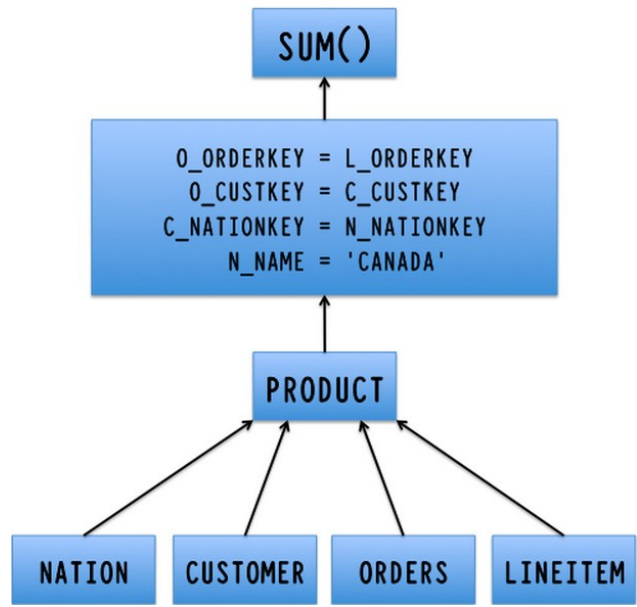

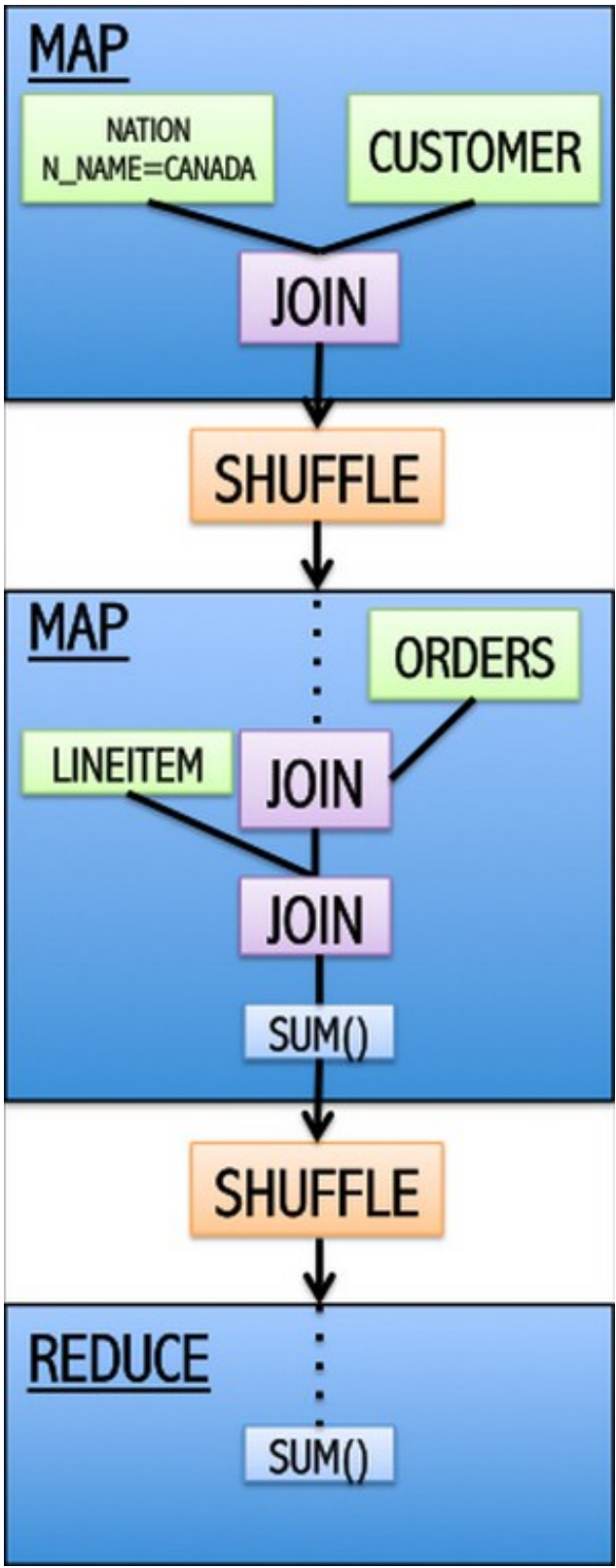

## VII. EXPERIMENTAL RESULTS

We have run micro benchmarks as well as the TPC-H benchmark at 100GB scale on clusters ranging from 4 to 32 nodes. All nodes are Amazon EC2 m3.2xlarge instances. These instances all have 8 cpus and 30GB of RAM. HRDBMS was compared against Hive and DB2.

**Micro-benchmarks**. First we performed some micro-benchmarks which show that HRDBMS is competitive with DB2 in all the key areas and greatly outperforms Hive.

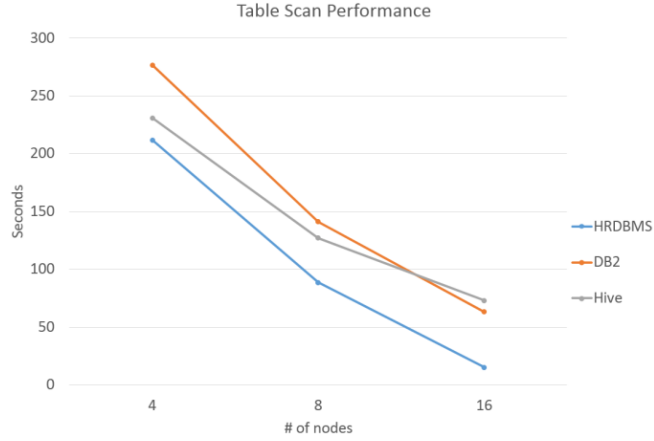

The table scan performance test measured how quickly the various databases could read a large table, apply a filter (where almost all rows pass), and return the data. HRDBMS has excellent table scan performance due to parallel I/O and LZ4 compression.

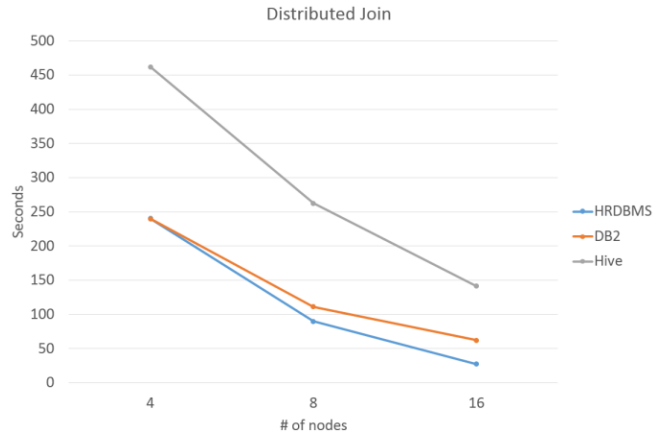

The distributed join test measured how quickly the various databases could perform a non-local join. That is, data first had to be shuffled before the join could be performed. Hive suffers here because of its blocking shuffle and unnecessary materializations.

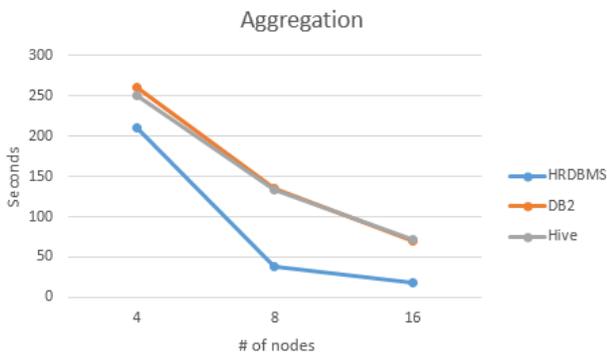

The aggregation test scanned a table, applied a filter (where most rows passed), and then performed aggregation. The filter was applied so that the aggregation could not be pre-computed and stored by the databases. Both Hive and DB2 are sorting before doing the aggregation (Hive has no choice). HRDBMS is instead doing hash aggregation, which we can see performs much better in this case.

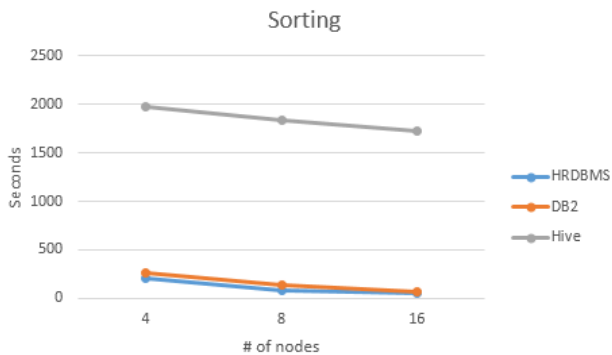

Unfrotunately, sorting in Hive is not handled very intelligently, and it's again due to limitations of the platform. MapReduce is capable of doing sorts very fast, but the results are placed into multiple HDFS files, which then have to be appended in the correct order. Since Hive needs the entire sorted set to be available to be returned to the client, it essentially just hands all the data to 1 node and sorts it (1 reducer). DB2 and HRDBMS correctly parallelize this sort across the cluster.

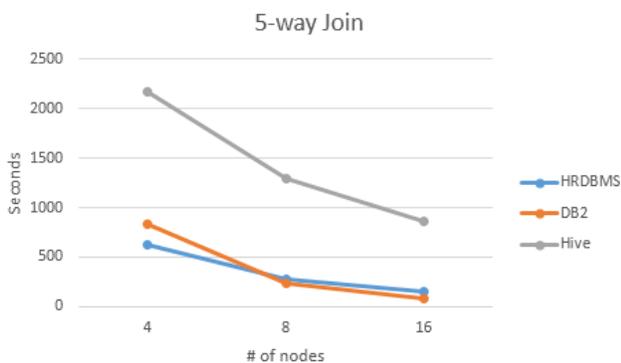

Joins of a large number of tables is notoriously difficult as the performance largely depends on the join order. It's good to see that the HRDBMS optimizer was up to the task and performed well compared to DB2. Hive, on the other hand, had to use a large number of MapReduce jobs to compute this result. This meant a lot of unnecessary materialization to disk.

**TPC-H**. Next we ran the TPC-H analytics benchmark on HRDBMS, DB2, and Hive. TPC-H is an industry standard benchmark for database performance when handling analytics queries. The data model consists of 8 tables: ORDERS, LINEITEM, CUSTOMER, PART, SUPPLIER, PARTSUPP, NATION, and REGION. The LINEITEM table is the bulk of the data. At the 100GB scale, the LINEITEM table contains approximately 600M rows. The benchmark then consists of 22 SQL queries that are designed to stress and expose weaknesses in database performance. The queries are very complex and include 1 query that has 2 levels of nested correlated sub-queries. A correlated sub-query is an inner query that is meant to be re-evaluated for every row in the outer query. The queries cover ever core SQL feature and do so in tricky ways that are likely to confuse optimizers, but are still very realistic as a type of analytics query that a business would be likely to run.

The results for TPC-H at the 100GB scale are shown below.

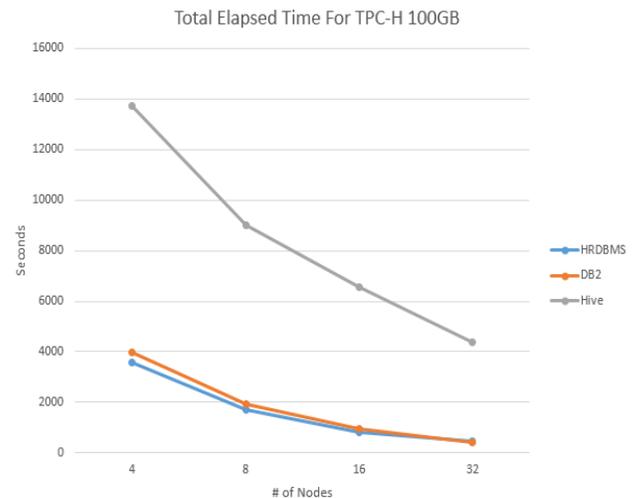

As you can see, both DB2 and HRDBMS were significantly faster than Hive. HRDBMS was slightly faster than DB2, but in general the performance was nearly equivalent. However, if we look at only DB2 and HRDBMS and look at performance by query we gain some more insight. Here is the performance per query at the 4 node scale.

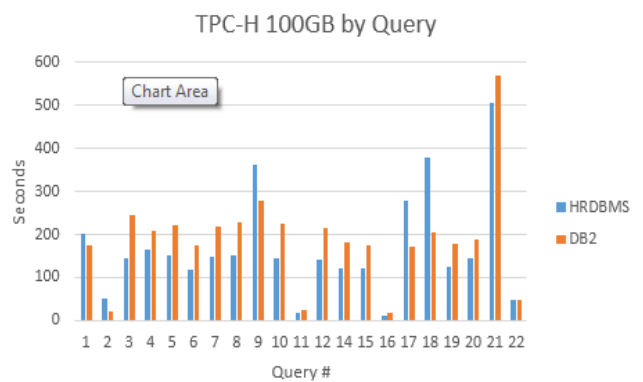

Here we see that for the majority of the queries, HRDBMS outperforms DB2 fairly significantly. On queries 1 and 2, HRDBMS is slightly slower. But, on queries 9, 17, and 18 HRDBMS performance is significantly worse than DB2. Queries 17 and 18 require further research on my part to see what is going on. On query #9, DB2 is taking a fairly non-standard approach. It is fetching only the columns that are available in the indexes, along with the associated RIDs. It is then applying all the predicates and the joins. At this point it has a smaller set of RIDs that have passed all conditions. It is then going back and fetching the necessary rows from all the tables and actually joining those rows together.

This type of approach would be very tricky to implement in an HRDBMS workflow. It would likely require 2 separate workflows with materialization to disk in between. I think that further testing will show that DB2's method to executing this query will not scale very well to larger clusters. I think that by restricting ourselves to what can be expressed as an HRDBMS worklow, the query will continue to scale well in HRDBMS.

If we go to the other end of the spectrum and look at per query performance at 32 nodes, we get these results.

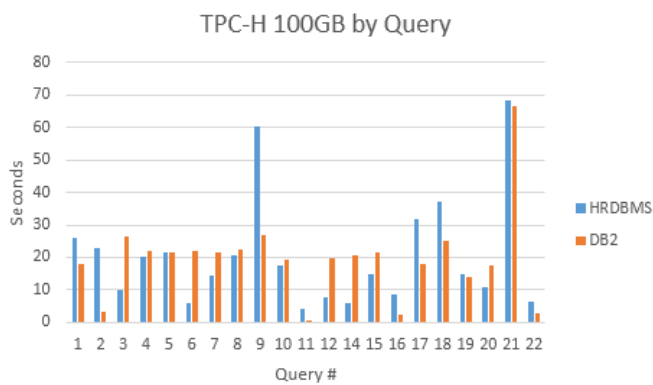

Now we see clearly that query #9 is the biggest problem, and there may be minor improvements I can make there, but again I believe that DB2 will start to scale poorly for query #9 when the cluster size is increased. Query #1 is still a little slower. Queries 17 and 18 still need to be addressed. Query 2 should be added to the list to look into further as well.

These results show that we were able to meet goal #1 of HRDBMS. We were able to create a MapReduce-like query execution engine capable of per node performance on par with a traditional MPP relational database. Further testing is required to validate goal #2 of being able to scale to larger clusters than a database like DB2.

## VIII. Future Work

Besides looking further at the performance of queries 2, 9, 17, and 18, there are many other items on my list for future work.

All testing so far has been done on clusters from 4 – 32 nodes in size. At these scales we still don't see the point where the traditional relational databases (like DB2) start to have scalability issues. Since the major premise of HRDBMS is to have on par per-node performance but scale further, we need to do some testing at these larger scales. I'm sure this will uncover bottlenecks in certain areas of HRDBMS that will need to be researched and addressed.

One of the challenges in testing with larger clusters is that we also need to use larger data sets. If we scale to clusters larger than 32 nodes using the 100GB data set, we will hit the point where data fits in memory. Traditional relational databases, such as DB2, excel in these types of situations and will out-perform HRDBMS by an order of magnitude. This is not the intention or focus of HRDBMS to be a database for small data. For doing analytics of big data on clusters, no one would ever even set up a cluster as big as 32 nodes just for 100GB of data, so I don't see this as a shortcoming of HRDBMS. However, it does imply that we don't want to scale larger than 320 nodes with the 1TB TPC-H data set, and so forth.

To test with these larger data sets, I will have to spend some time researching and improving the load performance of HRDBMS. Today it takes about 72 minutes to load the LINEITEM table (about 75% of the data) at 100GB and 4 nodes. The load scales well in that the same load only takes 20 minutes at 16 nodes. But, the overall load performance needs to be improved as taking 12 hours to load the LINEITEM table at 1TB and 4 nodes is way too slow. I believe this is a result of contention on locks used for synchronization (called latches in the DB world).

I am also doing some other research and novel work with loads. I have built a basic infrastructure that allows HRDBMS to sit on top of an HDFS cluster and then load the data from HDFS. By doing this, we allow all of the cluster nodes to be reading and parsing the data. If we don't load from HDFS, all of this work is placed on a single coordinator node. I still have a lot more research and work to do in this area before it is fully ready to go.

Additionally, statistics collection performance needs a lot of work. HRDBMS uses sampling to estimate the actual statistical values. But, this sample-based statistics collection is not scaling well today. It currently takes about 5 minutes at 4 nodes, but takes 20 minutes at 32 nodes. This needs to be investigated further.

I would also like to add additional databases to the testing. This should include at least Spark SQL if not others.

You'll notice that query 13 is not listed in the charts in the previous section. This is because HRDBMS does not yet have support for outer joins. The majority of basic SQL functionality is already built except for this. I consider this a major shortcoming that needs to be addressed.

Fault tolerance is not yet implemented in HRDBMS, but the system is already rack-aware and aware of how data is partitioned across nodes and disks. We plan to have HRDBMS maintain on-rack and off-rack replicas that can be used in the event of primary node failure. This could be extended to preferring secondary copies of data for load balancing. Data modifications affecting nodes that are down would be placed into pending-work queues. The work in these queues must be completed and successfully committed before a node is allowed to rejoin the cluster.

Lastly in the current HRDBMS architecture, the coordinator nodes can be a bottleneck. This is because all data has to eventually flow back through the coordinator node before being returned to the client. I'd like to do some research into removing this requirement. I envision that the various worker nodes could potentially open socket connections directly back to the client and the client could receive from multiple streams in parallel. Sorting could be handled by range partitioning the data across the workers and sorting before returning data to the client. As long as all the workers in the sort were given an ID representing their position in the overall range, the client could piece together the sorted data without even needing to do a mergesort. Likewise, the other final relational operations that are performed on the coordinator today could all be made to work in this type of manner. This now pushes the bottleneck all the way back to the client, and I think it will result in some significant improvements.

REFERENCES

[1] IBM. (2007, Oct 15). TPC Benchmark™ H Full Disclosure Report [Online]. Available: http://c970058.r58.cf2.rackcdn.com/fdr/tpch/IBM_570_10000GB_20071015_FDR.pdf

[2] IBM. (2014, Oct 17). IBM PureData System for Analytics N3001 [Online]. Available: https://www-304.ibm.com/connections/forums/html/topic?id=0e45d217-0912-442a-9033-a221495e9ccc

[3] J. Dietz. (2007 Sep). At the core of the Teradata platform [Online]. Available: http://apps.teradata.com//TDMO/v07n03/Tech2Tech/AppliedSolutions/TeradataPlatform.aspx

[4] EMC. (2002 Jul 1). Greenplum Database Architecture [Online]. Available: http://www.greenplumdba.com/greenplum-database-architecture

[5] J. Appleby. (2014, Dec 10). SAP HANA – Scale-up or Scale-out Hardware? [Online]. Available: https://blogs.saphana.com/2014/12/10/sap-hana-scale-scale-hardware/

[6] A. Thusoo, J. Sarma, N. Jane, et al, "Hive: a warehousing solution over a map-reduce framework,", PVLDB, Vol. 2, pp. 1626-1629, August 2009

[7] R. Xin, J. Rosen, M. Zaharia, et al, "Shark: SQL and rich analytics at scale,", SIGMOD, pp. 13-24, 2013

[8] D. Campbell, G. Kakivaya, N. Ellis, "Extreme scale with full SQL language support in microsoft SQL Azure,", SIGMOD, pp. 1021-1024, 2010

[9] S. Melnik, A. Gubarev, J. Long, et al, "Dremel: interactive analysis of web-scale datasets,", PVLDB, Vol. 3, pp. 330-339, September 2010

[10] M. Asay. (2014, Sep 12). Why the world's largest Hadoop installation may soon become the norm [Online]. Available: http://www.techrepublic.com/article/why-the-worlds-largest-hadoop-installation-may-soon-become-the-norm/

[11] J. Corbett, J. Dean, M. Epstein, et al, "Spanner: Google's Globally Distributed Database,", TOCS, Vol. 31, Issue 3, Article 8, August 2013

[12] J. Shute, R. Vingralek, B. Samwel, etal, "F1: a distributed SQL database that scales,", PVLDB, Vol 6, Issue 11, pp. 1068-1079, August 2013

[13] T. Li, X. Zhou, D. Zhao, etl al, "ZHT: A Light-Weight Reliable Persistent Dynamic Scalable Zero-Hop Distributed Hash Table,", IPDPS, pp. 775-787, 2013

[14] J. Standen. (2010, Feb 10). Inner and outer joins SQL examples and the Join block [Online]. Available: http://www.datamartist.com/sql-inner-join-left-outer-join-full-outer-join-examples-with-syntax-for-sql-server

[15] A. Wilschut, P. Apers, "Pipelining in Query Execution,", PARBASE, March 1990

[16] J. Mullin, "Optimal semijoins for distributed database systems,", Software Engineering, IEEE Transactions on, Vol. 16, Issue 5, pp. 558-560, May 1990

[17] G.M. Lohman, "Is Query Optimization a 'Solved' Problem?,", Workshop on Database Query Optimization (CSB Tech. Report 89-005), June 1989

[18] Collet, Yann. (2013). LZ4: Extremely fast compression algorithm [Online]. Available: code. google. com